\documentclass[prb,amsmath,aps,twocolumn]{revtex4}
\usepackage{graphicx,xspace}
\usepackage{float}
\usepackage{amsmath}
\usepackage{amssymb}
\usepackage[normalem]{ulem}
\input{epsf}
\begin{document}
\def\tit#1#2#3#4#5{{\sl #5} #1 {\bf #2} #3 (#4)}
\newcommand{\VEC}[1]{\mathbf{#1}} \newcommand{\TO}{,\ldots,}
\newcommand{\rvec}{\VEC{r}} \newcommand{\scalar}[2]{\left \langle#1\
    #2\right \rangle} \newcommand{\me}{\mathrm{e}}
\newcommand{\mi}{\mathrm{i}} \newcommand{\dif}{\mathrm{d}}
\newcommand{\ket}{\rangle}
\newcommand{\fcs}{f_{\text{c}}^{\text{smpl}}}
\newcommand{\bra}{\langle} \newcommand{\half}{\frac{1}{2}}
\newcommand{\mean}[1]{\langle #1 \rangle}
\newcommand{\fig}[1]{figure~\ref{#1}}
\newcommand{\msmall}{\scriptscriptstyle}
\newcommand{\eq}[1]{eq.~\ref{#1}} \newcommand{\gl}{\!=\!}
\newcommand{\mn}{\!-\!}  \newcommand{\nufs}{\nu_{\msmall \text{FS}}}
\newcommand{\pf}{\!\rightarrow\!}
\newcommand{\Dfix}{\Delta^*}
\newcommand{\mq}[2]{\uwave{#1}\marginpar{#2}} 
\newcommand{\gae}{\lower 2pt \hbox{$\, \buildrel {\scriptstyle >}\over
    {\scriptstyle \sim}\,$}} \newcommand{\lae}{\lower 2pt \hbox{$\,
    \buildrel {\scriptstyle <}\over {\scriptstyle \sim}\,$}}

\title{Chaos in the thermal regime for pinned manifolds via functional RG}
\author{Olaf Duemmer} \email{duemmer@lps.ens.fr}
\affiliation{CNRS-Laboratoire de Physique Statistique,
  Ecole Normale Sup{\'{e}}rieure, 
  24 rue Lhomond, 75231 Paris Cedex 05, France}
\author{Pierre Le~Doussal} \email{ledou@lpt.ens.fr}
\affiliation{CNRS-Laboratoire de Physique Th\'eorique, 
  Ecole Normale Sup{\'{e}}rieure,
  24 rue Lhomond, 75231 Paris Cedex 05, France}
\begin{abstract}
The statistical correlations of two copies of a $d$-dimensional elastic manifold
embedded in slightly different frozen disorder are studied using
the Functional Renormalization Group to one-loop, order $O(\epsilon=4-d)$,
accuracy. 
Determining the initial (short
scale) growth of mutual correlations, i.e. chaos exponents, 
requires control of a system of coupled differential (FRG) equations
(for the renormalized mutual and self disorder correlators) in a
very delicate boundary layer regime. Some progress is achieved 
at non-zero temperature $T>0$, where linear analysis can be used.
A growth exponent $a$ is defined from center of mass fluctuations
in a quadratic potential. In the case where temperature is
marginal, e.g. a periodic manifold in $d=2$, we demonstrate analytically and
numerically that $a=\epsilon (1/3 -1/(2 \ln(1/T))$ with
interesting and unexpected logarithmic corrections at low $T$. 
For short range (random bond) disorder our analysis indicates
that $a= 0.083346(6) \epsilon$ with large finite size corrections.
\end{abstract}

\maketitle

\section{Introduction}

\subsection{Overview}

Systems with quenched disorder are especially sensitive to small
external perturbations. This phenomenon-is called chaos.
Upon a perturbation of amplitude $\delta$ (an energy scale), e.g. a
small change in temperature (temperature chaos) or in the
disorder (disorder chaos) the configuration of the system, while
only weakly affected at small scales, changes completely beyond a
length scale $L_\delta$. This overlap length
diverges as $L_\delta \sim \delta^{-\alpha}$ for small $\delta$,
$\alpha$ being called the chaos exponent. Chaos can be studied
either at $T=0$ via the sensitivity of the ground state to
perturbations, or in the thermal regime at $T>0$. 
Chaos has been studied in spin glasses and other
disordered systems using droplet arguments
\cite{mckay,braymoore,droplets}, numerics \cite{rieger} and mean field calculations
\cite{mf} . While
central to disordered systems, chaos is still not fully understood.
There has been some controversy as to whether the overlap length is
finite or infinite (no chaos), in high dimensions as well as in mean field
\cite{nifle,mf}.

An interesting class of systems exhibiting chaos are elastic systems
in random potentials such as domain walls in
disordered magnets \cite{creepexp}, or periodic systems such as
charge density waves \cite{cdw} and vortex lattices on disordered
substrates \cite{vortices,braggglass}. They have energy-dominated
glass fixed points at which temperature is irrelevant, $T_L \sim L^{-
\theta}$, $\theta = d - 2 + 2 \zeta$ being the free energy fluctuation
exponent. The roughness exponent $\zeta$ controls the scaling  
of the typical deformation $u \sim L^\zeta$ with 
the internal size $L$  of the pinned
configurations of the elastic object. Chaos in pinned
manifolds was studied mostly via scaling arguments \cite{shapir}.
The directed polymer ($d=1$) was studied numerically and via
analytical arguments for $N=1$ indicating $\alpha=1/6$ in agreement
with droplets \cite{Zhang}, and recently, on hierarchical lattices
\cite{silveira}. In $d=2$ chaos was demonstrated for periodic
systems near the glass transition $T_g$ using the Cardy Ostlund RG
\cite{hwafisher}.

A successful approach to disordered elastic systems is the
functional renormalization group (FRG). It allows an efficient determination
of the roughness of the ground state within an
expansion in $\epsilon=4-d$, to one loop \cite{fisher,frgdep,bg} and
recently to higher orders \cite{twoloopdep}. FRG involves a coupling
constant function, $\Delta(u)$ which measures the renormalized
correlator of the pinning force and becomes non-analytic beyond the
(Larkin) length scale $L_c$ where pinning produces metastability.
FRG was also studied at non-zero temperature to one loop
\cite{ChauveBL}. There it was shown \cite{balentspld} that rare
thermal fluctuations (droplets) lead to the cusp of $\Delta(u)$ being rounded off
inside a region of size $u \sim T$, the so-called thermal
boundary layer (TBL).
Recently \cite{pld} the full function
$\Delta(u)$ was shown to be a proper physical observable, 
which describes the fluctuations of the center of mass
of an elastic manifold confined by an harmonic well. It was determined by a high precision
numerical calculation \cite{alan} at $T=0$
and found to compare remarkably well with analytical predictions, 
already at the one loop level. 
The cusp in $\Delta(u)$ was also observed \cite{alan}, and shown to result from 
so called shocks,
abrupt switches of the manifold from one ground state to another, as the
position of the center of the well is varied.

When applying FRG to the problem of chaos
\cite{PierreChaos}, one must follow not only the flow of $\Delta(u)$, but also the flow of 
a second FRG correlation function $D(u)$ which encodes the mutual
correlations between the centers of mass of two manifold copies seeing slightly
different disorders. These flows have been analyzed mostly for large scale mutual correlations. 
It was found \cite{PierreChaos} that the residual correlations decay to zero at large scale
for (i) the 'random periodic' class (RP),
i.e. a correlator $\Delta(u)$ periodic in $u$ (which describes charge density waves
and vortex lattices) and for (ii) the 'random bond' class (RB), i.e. a
correlator $\Delta(u)=-R''(u)$ where $R(u)$ is a short range
function (which describes magnetic domain walls in short range
disorder).
Hence there is chaos with a finite $L_\delta$. For the 'random
field' class (RF), which describes magnetic interfaces in the presence of
random fields, it was found that residual correlations remain non-zero. 
However, determining the {\it initial short scale} growth of
mutual correlations, and hence the chaos exponents, is difficult. 
It requires good control of the system of coupled
differential one-loop FRG equations, specifically of the separation---initially very small---
between $D(u)$ and $\Delta(u)$. Since
$D(u)$ remains analytic (which was confirmed numerically \cite{alan})
while $\Delta(u)$ develops a cusp at $u=0$,
one cannot use standard linear analysis. Instead, the chaos boundary
layer, where the two functions differ, must be investigated: a
non-trivial task in full generality.

The aim of the present paper is more modest. We study the chaos
problem on the one-loop FRG equation in the thermal regime $T >0$, where linear
analysis is applicable. The easiest case to handle
is when temperature does not flow under RG and there is a line of
fixed points as temperature is varied. This happens
for the random periodic class and $d=2$, the case on which we focus here. Hence for
each $T>0$ there is an analytic FP, $\Delta_T(u)$, and one can use
linear analysis to extract the growth exponent for mutual
correlations. It is still a non-trivial task as one must perform the
analysis both inside and outside the TBL, and match the two results. 
Surprisingly, as $\Delta_T(u)$ becomes non-analytic for $T\to 0$, one finds
logarithmic corrections to the growth exponent. These corrections are
confirmed by a careful numerical study of the differential equation.
Despite being a special case of the full chaos problem, the random periodic 
class for $d=2$ already illustrates the difficulty of obtaining the accurate
behavior. An extension to the random bond class is then
proposed, again within the thermal regime.

\subsection{The model and the observables}

We investigate how two identical copies of an harmonic elastic
manifold embedded in frozen disorder decorrelate when they are
exposed to slightly different disorder. Here we focus on interfaces,
i.e. manifold whose deformations are parameterized by a real valued
displacement field $u(x)$, where $x$ is the $d$-dimensional internal
coordinate. The system is described by the following Hamiltonian
\begin{eqnarray}
H_{V,v}[u] &=& \sum_{i=1}^2 \int \dif^d x \bigg[\frac 1 2 (\nabla u^i)^2 \\
  && {} + V_i(u^i(x),x) + \frac{1}{2} m^2 (u^i(x) - v)^2 \bigg] \nonumber
\end{eqnarray}
where the two copies $i=1,2$ are not mutually interacting. They are
however coupled via the correlations between the two random
potentials, and between the corresponding random pinning forces
$F_i(u,x) = - \partial_u V_i(u,x)$, whose correlation matrices take
the form:
\begin{eqnarray}
&& \overline{V_i(u,x) V_j(u',x')} = R^{(0)}_{ij}(u-u')
\delta^d(x-x')  \label{correlator} \\
&& \overline{F_i(u,x)F_j(u',x')} = \Delta^{(0)}_{ij}(u-u')
\delta^d(x-x').
\end{eqnarray}
with $\Delta^{(0)}_{ij}(u)=- R^{(0)''}_{ij}(u)$. We use the
superscript $(0)$ to denote bare disorder, to distinguish it from
the renormalized disorder defined below. The form of these correlation functions 
differentiates the three main universality classes: (i)
random periodic (RP): a periodic $R^{(0)}(u)$ (ii) random bond (RB)
a short range function $R^{(0)}(u)$ (iii) random field (RF)
$R^{(0)}(u) \sim \sigma |u|$ and $\Delta^{(0)}(u)$ short range. One
way of realizing eq.~(\ref{correlator}) is to consider two disorder copies
of the following form: $V_i(u,x) = V(u,x) \pm \delta
W(u,x)$ with $V$ and $W$ mutually uncorrelated Gaussian
disorders of zero mean, and $\delta$ a very small
parameter. In the simplest case $V$ and $W$ are in the same class with identical
correlator. Note that a quadratic well, i.e. a mass term, has been added
to confine each elastic manifold copy to a mean position $v$, another external parameter.
Its role is discussed below. One is usually interested in the limit $m \to 0$.

The statistical correlations of the two ground states
are measured by the standard observables, i.e. the
correlation functions:
\begin{equation} 
\label{corr}
C_{ij}(x-x') = \overline{ \bra (u^i(x) - u^i(x'))\ket  \bra  (u^j(x)-u^j(x')) \ket }
\end{equation}
where $\overline{..}$ denotes disorder averages. We
will also study non-zero, though low, temperature, denoting the thermal averages 
 by $\bra .. \ket$ (using the canonical
partition function $Z_{V,v}= \int Du e^{- H_{V,v}[u]/T}$). For a single copy one expects
mean square deformations due to disorder to result in $C_{11}(x) \sim
c_{11} |x|^{2 \zeta}$ at large $L_c \ll x \ll \frac{1}{m}$, $\zeta$ being the roughness
exponent, independent of temperature at low $T$ (and in the whole
glass phase where the manifold is pinned near its ground state with
only a few active thermal excitations). Standard arguments that assume
the existence of a single diverging scale, the overlap length
$L_{\delta} \sim \delta^{-1/\alpha}$, suggest the following
scaling form for the two-point correlation function between
different copies at large $L_c \ll x \ll \frac{1}{m}$:
\begin{equation}
C_{12}(x) = x^{2\zeta} \Phi(\delta x^{\alpha}). \label{chaoscorr}
\end{equation}
The overlap length separates correlated from uncorrelated scales,
and depends sensitively on the small difference $\delta$ between the
bare disorders of the two copies. In analogy with chaotic dynamical
systems, in which tiny differences in initial conditions are
amplified via the Lyapunov exponent(s) to large scale differences,
one introduces the {\em chaos} exponent $\alpha$. This exponent is
a measure of how the two copies effectively split as
scale increases. Qualitatively, this splitting is characterized
by a dimensionless scale-dependent parameter which grows under RG as 
$\tilde \delta_l = \delta e^{\alpha l}$ where $l$ is the log-scale, e.g.
$l=\ln L$.
This behavior is suggested by droplet arguments, first developed
for spin glasses where they predict \cite{braymoore} 
$\alpha=d_f/2 - \theta$ with $-\theta$ being the thermal eigenvalue
and $d_f$ the fractal dimension of the droplets. In the case of
manifolds the same formula was proposed \cite{droplets,shapir} for
the SR disorder class, with $d_f=d$, namely $\alpha=d/2 - \theta=(\epsilon - 4 \zeta)/2$.

Another observable, introduced recently \cite{pld}, quantifies the fluctuations
of the center of mass confined by an harmonic well. One defines
$u^i(x;v) = \langle u^i(x) \rangle$ the thermally averaged position.
It depends on the position of the center of the harmonic well
$v$. One denotes the center of mass of the manifold
by $\bar u^i(v) = L^{-d} \int d^d x
u^i(x;v)$, $L^d$ being the system volume. The second cumulant of its position, as the disorder is
varied:
\begin{equation} \label{deltadef}
m^4 \overline{ (\bar u^i(v) - v) (\bar u^j(v') - v')} = L^{-d}  \Delta_{ij}(v-v')
\end{equation}
defines the renormalized pinning force cross-correlator $\Delta_{ij}(u)$.
At zero temperature these functions measure the correlations between the
shocks in the two copies. These abrupt jumps of the manifold as the center $v$
is varied do not occur exactly at the same place in the two copies, which results
in the cross correlator $\Delta_{12}(u)$ remaining a smooth function of $u$.
This was confirmed in \cite{alan} where simultaneous shocks in the two copies
were examined and $\Delta_{12}(u)$ was computed numerically.
Another useful quantity is the free energy of each copy
$\hat V_i(v) = - T \ln Z_{V,v}$: for each disorder configuration it
is a random function of the well center position $v$, hence one
defines the renormalized second cumulant of the potential:
\begin{equation} \label{rdef}
\overline{ \hat V_i(v) \hat V_j(v') } = L^d R_{ij}(v-v')
\end{equation}
and it is easy to see that $\Delta_{ij}(u) = - R''_{ij}(u)$.

\subsection{Functional RG approach}

The functional RG method allows us to compute from first principles the observables defined above, namely the correlation functions of eq.~(\ref{corr}) and the renormalized correlators of eq.~(\ref{deltadef}) and (\ref{rdef}). FRG is based on the replicated field theory, and proceeds via a loop expansion for which --- at zero temperature --- the small parameter is $\epsilon=4-d$. Here we give only results, and refer to \cite{frg_review,pld} for reviews on the method. The FRG flow of the renormalized correlators $R_{ij}$ were derived in \cite{PierreChaos}. The FRG flow equations for the correlators $R_{ij}$ and $\Delta_{ij}$, defined in eq.~(\ref{deltadef}) and (\ref{rdef}), are found by computing the effective action and its flow equation, as the mass is varied. One
defines $m \equiv m_l = m_0 \mathrm{e}^{-l}$ where $l$ is the usual
RG logarithmic scale \endnote{Strictly speaking, for the observable of eq.~(\ref{deltadef})
one has the initial condition of the flow $\Delta(u)=\Delta^0(u)$ for
$m=+\infty$, i.e. $l=-\infty$.}. We define the rescaled
dimensionless force correlator $\tilde{\Delta}_{ij}$ via $\Delta_{ij}(u) = A_d^{-1}
m_l^{\epsilon - 2 \zeta} \tilde{\Delta}_{ij}(u m_l^{\zeta})$, with
$A_4^{-1} = 8\pi^2 $, and the roughness exponent $\zeta$
reflecting the self-affine scaling property of the manifold. A
rescaled temperature is also defined as $T_l=T m_l^\theta$. We
simplify notation by writing $\Delta \equiv \tilde{\Delta}_{11}$ for
the one-copy correlator and $D \equiv \tilde{\Delta}_{12}$ for the
two-copy correlator. Their flow equations are \cite{PierreChaos},
respectively, to one loop:
\begin{subequations}
\label{eq_delta_flow}
\begin{eqnarray}
\partial_l \Delta(u) &=& (\epsilon-2\zeta)\Delta(u) + \zeta u \Delta'(u) - \Delta'(u)^2 \\
 && {} + \left[ \Delta(0) - \Delta(u) \right ] \Delta''(u) + T_l \Delta''(u) \nonumber \\
\partial_l D(u) &=& (\epsilon-2\zeta)D(u) + \zeta u D'(u) - D'(u)^2 \\ 
 && {} + \left[  \Delta(0) - D(u) \right] D''(u) + T_l D''(u)  \nonumber
\end{eqnarray}
\end{subequations}
The zero temperature FRG equation
are obtained by setting $T_l=0$.

Note  that the temperature is irrelevant (for $\theta>0$) 
because $T_l= T e^{-\theta l}$ (setting $m_0=1$). However, the $T_l \Delta''(u)$ term keeps 
the correlation functions smooth for any non-zero $T$.
If one studies the FRG to only one loop, one can
also use the (more qualitative) Wilson RG procedure, which consists of varying the short scale momentum cutoff $\Lambda_l = \Lambda_0 \mathrm{e}^{-l}$. In that case the mass cutoff is unnecessary: one can set $m=0$ and estimate the correlation functions of eq.~(\ref{corr}) at non-zero momentum.

It is important to note that the one-copy correlator in eq.~(\ref{eq_delta_flow})
evolves independently, whereas the two-copy correlator is linked to
the former via its value at the origin $\Delta(0)$. This small but
crucial difference entails opposing behaviors for the two
correlators \cite{subdominant}: the one-copy correlator converges to its stable fix
point, whereas the two-copy correlator diverges towards another
fixed point.  The difference between the two correlators is denoted
\begin{eqnarray}
\label{eq_that}
 & \Theta_l & \equiv \Delta(0) - D(0) \\ 
 & \Theta_{l=0} & \delta^d(x-x') = \nonumber \\
 && \frac 1 2 \overline{(F_1(0,x)-F_2(0,x)) (F_1(0,x')-F_2(0,x'))} \nonumber
\end{eqnarray}
One easily sees from eq.~(\ref{eq_delta_flow}) that $\Theta_l$
generates an additional term $\Theta_l D''(u)$
in the flow of $D(u)$, compared
to that of $\Delta(u)$. Hence one can think of $\Theta_l$ as an 
effective temperature (its bare value before renormalization is
$\Theta_{l=0} \sim \delta^2$) but its flow under RG is very
different from that of the real temperature $T$. In fact, its flow is
determined self-consistently by the two equations. While in general
it smoothes the form of $D(u)$ which hence remains analytic \cite{subdominant}
, it is often relevant, i.e. grows with $l$, by contrast to temperature.

To compute the observables of eq.~(\ref{deltadef}) and (\ref{rdef}) as a function of $m$ one must solve the above flow equations. The difference (eq.~(\ref{eq_that})) is a direct measure of the fluctuations of the distance between the (thermally averaged) centers of mass $\bar u^i(v)$ of the
two copies:
\begin{equation} \label{diff}
\overline{ (\bar u^1(v) - \bar u^2(v))^2 } = A_d L^{-d} m_l^{-d-2 \zeta} \Theta_l
\end{equation}

The correlation function of eq.~(\ref{corr}) is more delicate to compute.
The general formula $C_{ij}(q)= \Delta_{ij,l}(0)/(q^2+m^2)^2$
is exact for $q=0$, $m=m_l$, and holds to $O(\epsilon)$ (i.e.
to one loop accuracy) for $q \sim m$. It also holds for
$q=\Lambda_l$, $m=0$ within the (one loop) Wilson scheme
and hence provides an estimate for the correlation function
at large scale (and for $q \gg m$):
\begin{equation}
C_{ij}(q) \approx A_d q^{-4} [ \tilde{\Delta}_{l,ij}(0) e^{(2 \zeta
- \epsilon) l} ]_{l=\ln(\frac{\Lambda_0}{q})}
\label{general}
\end{equation}
However caution is required when using this estimate, even when computing the simpler, small
$q$ behavior. For instance, for the random periodic (RP) universality class (for which $\zeta=0$) 
the estimate is correct only for $d>2$. In general one must examine more carefully
the FRG for the non-local part of the effective action \cite{SchehrCO}: in
$d=2$, $\theta=0$ the two gradient term 
becomes dominant and yields an extra $\ln(1/q)$ in the single copy
correlation and the famous $\ln^2|x|$ in the
real space correlation \footnote{As was shown there, such a term can be generated at $T=0$
only from the non-analyticity in the disorder correlator. Since the
two copy disorder correlator $R_{12}$ remains analytic, such a term should not be generated in the two copy mutual correlation $C_{12}$ which hence should remain a single logarithm.
This is consistent with numerical results (G. Schehr, H. Rieger unpublished).}.
The study needed to elucidate the initial growth regime being even more
subtle, our results here will mostly concern the $q=0$, center-of-mass
behavior of eq.~(\ref{diff}) and (\ref{deltadef}). 

Finally note that the thermal correlations:
\begin{eqnarray} \label{corr_thermal}
\lefteqn{C^{th}_{ij}(x-x') = } \\
 && \quad \overline{ \bra (u^i(x) - u^i(x')) (u^j(x)-u^j(x')) \ket} \nonumber \\
 && {} - \overline{ \bra (u^i(x) - u^i(x'))\ket  \bra  (u^j(x)-u^j(x')) \ket } \nonumber
\end{eqnarray}
always vanish identically for $i \neq j$, because the two copies do not interact \footnote{This agrees with a droplet estimate in the case of two almost degenerate wells in each copy, each thermally occupied with probabilities $p_i$ and $1-p_i$ respectively, provided the probabilities of joint thermal occupations are products
of the form $p_1 p_2$. It is obvious that in a given disorder sample $(V_1,V_2)$ thermal occupations are statistically independent in each copy
if there are no interactions between the copies. However, there are obviously correlations between the random variables $p_1$ and $p_2$ with respect to the measure on the disorder.}.

The outline of the paper is as follows. We start by studying the simplest case of the random periodic class. In section \ref{RPT0} we focus on zero temperature, review known results and explain why the problem is difficult. In section \ref{RPT} we study the RP problem at $T>0$ in the so called marginal case of $\theta=0$ (i.e. in $d=2$) where temperature does not flow. This allows to use linear analysis. In section \ref{sec:num} analytical predictions and numerical analysis are compared. In section \ref{all} we generalize these results to the random bond class, and RP for $d=4-\epsilon$.
The results are summarized and discussed in the conclusion.

\section{Random periodic class: zero temperature considerations}
\label{RPT0}

\subsection{Flow of single copy and fixed point of the random periodic problem (CDW, Bragg glass)}

Here we study the random periodic class (RP), which has logarithmic roughness, i.e $\zeta = 0$. The single-copy correlator is a periodic function $\Delta(u+1) = \Delta(u)$ of normalized period one, and obeys the following flow equation at $T=0$, from eq.~(\ref{eq_delta_flow}):
\begin{eqnarray}
\partial_l \Delta(u) &=& \epsilon \Delta(u) - \Delta'(u)^2 \\
 && {} + (\Delta(0) - \Delta(u))\Delta''(u) \nonumber
\end{eqnarray}
where $\epsilon=4-d$. This flow is well understood \cite{fisher,bg,ChauveBL}. Beyond the Larkin scale --- here $m=m_c=m_0 e^{l_c}$ --- it develops a non-analyticity (cusp) at $u=0$, and flows towards an attractive fixed point:
\begin{eqnarray}
  \label{eq_fix}
  \Dfix(u) &=& \frac {\epsilon} {36} (1  - 6 u(1-u)) \quad , \quad u \in [0,1]
\end{eqnarray}
Note that this correlator presents a cusp at $u=0$ because of the periodicity condition $\Delta(u+1)=\Delta(u)$.

\subsection{Flow of the two-copy correlator}

The flow of the two-copy correlator is more intricate, due to the (scale dependent) coupling $\Theta_l$ to the single-copy correlator.
\begin{eqnarray}
  \partial_l D(u) &=& \epsilon D(u) - D'(u)^2 \\ 
 && {} + (\Theta_l + D(0) - D(u))D''(u) \nonumber \\
  \label{eq_two_copy_flow}
                  &=& \epsilon D(u) - D'(u)^2 \nonumber \\ 
 && + (\Delta(0) - D(u))D''(u) \nonumber
\end{eqnarray}

The only difference
between the two equations is the term $\Theta_l D''(u)$. If one starts with a very small $\Theta_0 \sim \delta^2 \ll 1$, the two correlators $\Delta(u)$ and $D(u)$ remain practically identical up to the Larkin scale $l_c$, and the difference $\Theta_l = \Delta(0) - D(0)$ remains small. One finds that it grows as $\Theta_l \sim \mathrm{e}^{2 \tilde a_L l} \Theta_0$, with $\tilde a_L = (\epsilon - 2 \zeta)/2$, since in most of this regime one can neglect the non-linearities. Near the Larkin scale, non-linearities become important and $\Delta''$ becomes large as the cusp develops. Once $\Theta_l$ grows such that $\Theta_l D''(0) \sim \Theta_l \Delta''(0) \sim \epsilon D(0)$, which occurs very near $l_c$, the two-copy correlator $D(u)$ starts to differ from the one-copy correlator. By analogy with temperature, one expects this difference to be mostly confined to a boundary layer (BL) of width $u \sim \Theta_l \ll 1$ around $u=0$. This BL is called chaos BL to distinguish it from the thermal BL $u \sim T_l$ (absent at zero temperature).
While the one-copy correlator flows towards its fixed point and develops a cusp (eq.~(\ref{eq_fix})), the two-copy correlator remains analytic.

The flow beyond the Larkin scale is non-trivial. In Ref. \cite{PierreChaos} the final behavior of the flow for $l \to \infty$ was examined. It was found that ultimately
$D(u)$ flows to $D(u)=0$ for the RP class, hence there are no residual correlations between the two copies. Here we address a different question. We are interested in the
first phase of the FRG flow, i.e. we study how the difference $\Theta_l$ grows beyond the Larkin scale.

Clearly $\Dfix$ is a fixed point both for $D(u)$ and $\Delta(u)$. However while $\Delta$ flows towards its attractive fixed point $\Dfix$, $D$ is repelled by it. Let us assume that the one-copy correlator has already reached its fixed point $\Delta(u) = \Dfix(u)$ (eq.~(\ref{eq_fix})). We then have to solve the FRG equations for the flow of the two-copy correlator $D$, from which we can deduce the behavior of the difference $\Theta_l=\Dfix(0)-D(0)$ :
\begin{eqnarray}
\label{eq_flow_T0}
\partial_l D(u) &=& \epsilon D(u) - D'(u)^2 \nonumber  \\ 
 && {} + (\Dfix(0) - D(u))D''(u) \\
\partial_l \Theta_l &=& - \partial_l D(0) \nonumber 
\end{eqnarray}
One can always write:
\begin{equation}
D(u) = \Dfix(u) + f(u,l)
\end{equation}
where, during the initial growth phase, $f(u,l)$ remains small (in a sense to be made precise below). 
The problem we face at $T=0$, is that the one-copy fixed point correlator is non-analytic, whereas the two-copy correlator is analytic, with the cusp rounded off inside the chaos boundary layer. Hence the
function $f(u,l)$ should be equally non-analytic, to cancel the fixed point cusp and leave a smooth analytic function $D(u)$.

An important property of $f(u,l)$ arises from the potentiality constraint.
$\Delta_{ij}(u) = - R''_{ij}(u)$ can at most have a cusp singularity \footnote{That is $R'(0^\pm)=0$, i.e. no supercusp.} and for the RP class the $R_{ij}$ are periodic, while for the RB class they must be short ranged, which implies:
\begin{equation} \label{pot}
\int du f(u,l) = 0
\end{equation}
for both the RP and the RB class (the integration domains being $u \in [0,1]$ and $u \in [0,\infty]$, respectively.)

We now attempt a linear expansion around the fixed point. We need
eigenfunctions $f(u,l)$ that are non-analytic and that obey the zero-mean constraint of eq.~(\ref{pot}).
Although linear analysis is not necessarily valid inside the
boundary layer, outside (for $u \gg \Theta_l$) it is appropriate.

\subsection{Eigenvalue problem of the $T=0$ linearized flow equation}
\label{sec:zeroT}

One starts from:
\begin{eqnarray}
\Dfix(0) -  D(u) &=& \Dfix(0) - \Dfix(u) + f(u,l) \\
&=& \frac{\epsilon}{6} u(1-u)  + f(u,l)
\end{eqnarray}
and inserts it into the flow equation (\ref{eq_flow_T0}) keeping linear terms only. We assume the eigenvector flows as:
\begin{eqnarray}
f(u,l) = \exp(2al) f_a(u) 
\end{eqnarray}
which provides one definition of a growth exponent $a$, as discussed below.
One gets:
\begin{eqnarray}
\label{eq_flow_lin}
0 &=& (\epsilon-2a)f(u) + \frac {\epsilon} 6 [u(1-u)f(u)]'' \\
  &=& 4(1-\frac{3 a}{\epsilon})f(u) \nonumber \\ 
  && {} + 2 (1-2u)f'(u) + u(1-u)f''(u) \nonumber
\end{eqnarray}
For any value of $a$ this linear second order differential equation has two types of solutions 
 on the interval $u \in [0,1]$, even and odd about $u=1/2$. We must select the even one (since the correlators are symmetric: $\Delta(-u)=\Delta(u)$
and $D(-u)=D(u)$ combined with periodicity). The even solution for general eigenvalue $a$ can be expressed in terms of the hypergeometric function ${}_2 F_1$
\begin{eqnarray}
\label{eq_evector}
f_a(u) = & {}_2F_1 \bigg(& \frac 3 4 - \frac 1 4 \sqrt{25 - 48 \frac a{\epsilon}}, \\ 
 && \frac 3 4 + \frac 1 4 \sqrt{25-48\frac a{\epsilon}}; \frac 1 2; (1-2u)^2 \bigg) \nonumber
\end{eqnarray}
where ${}_2 F_1 (\alpha,\beta,\gamma,z) = \sum_{n=0}^\infty
\frac{ (\alpha)_n (\beta)_n }{(\gamma)_n} \frac{z^n}{n!}$, with $(\alpha)_n =
\Gamma(\alpha+n)/\Gamma(\alpha)$, being Gauss' hypergeometric series. It provides a convergent series for $0 < u < 1$, whose behavior near $u=0$ is:
\begin{eqnarray}
\label{expansion}
 f_a(u) &=& A_a \left[ \frac{1}{u} + B_a \ln u (1 +
b_1(a) u + b_2(a) u^2 +..) \right] \nonumber \\
 &&  {}+ c_0(a) + c_1(a) u + c_2(a) u^2 +..
\end{eqnarray}
with $B_a=12 \frac a \epsilon - 6$ and 
\begin{equation}
A_a = \frac{\sqrt{\pi} }{
4
\Gamma[\frac{3}{4} - \frac 1 4 \sqrt{25 - 48 \frac a{\epsilon}} ]
\Gamma[\frac{3}{4} + \frac 1 4 \sqrt{25 - 48 \frac a{\epsilon}} ] }
\end{equation}
Hence for generic $a$ there is a non-integrable divergence at $u=0$ together with terms non-analytic in $|u|$. Note a particularly simple solution for
$a=\epsilon/2$: $f_a(u) = 1/(u(1-u))$. It is however not integrable at $u=0$.

For $\int_0^1 f_a(u)$ to be defined (i.e. finite) one needs $A_a=0$, which gives:
\begin{equation}
\frac{3}{4} \pm \frac 1 4 \sqrt{25 - 48 \frac a{\epsilon}} = - n
\end{equation}
where $n$ is a positive integer. It yields a series of values:
\begin{equation}
a = a_n = \left( \frac{1}{3} - \frac{1}{3} n^2 - \frac{1}{2} n\right) \epsilon \quad , \quad n=0,1,2,..
\end{equation}
for which the hypergeometric series becomes a polynomial of finite order. For the highest value:
\begin{equation}
a = \frac{\epsilon}{3} \quad , \quad f_a(u) = c_0
\end{equation}
the eigenfunction is a constant. The next one is $a= - \frac{\epsilon}{2}$ and corresponds to $f_a(u) = c_e(1-5u(1-u))$. A more detailed analysis is performed in 
appendix \ref{appendix1}. It is found that none of these eigenfunctions satisfy the potentiality condition of eq.~(\ref{pot}) $\int_0^1 f_a(u)=0$.

This dilemma of non-zero mean eigenfunctions will be overcome in the next section, by studying the FRG flow at non-zero temperature. Setting $T>0$ induces the cusp to round off within a boundary layer around $u=0$, and permits solutions of the eigenvalue problem that have zero mean value.

More generally, the above analysis is valid outside the boundary layer (BL), be it a thermal BL $u \sim T_l$ or the chaos BL $u \sim \Theta_l$. One expects the blow-up of the eigenfunction near the origin to be rounded off within the BL. We now examine how.

\section{Random periodic universality class: non-zero temperature $T>0$
in the marginal case ($\theta=0$, $d=2$).}
\label{RPT}

To escape from the difficulty of a non-analytic fixed point $\Delta^*$ we now consider the problem at non-zero temperature. In this section we focus on the simplest case, $\theta=d-2=0$, where temperature does not flow under RG. Hence there is a line of analytic fixed points $\Dfix_T$, indexed by $T>0$, which converge to $\Delta^*=\Dfix_{T=0}$ at the end of the line $T \to 0^+$. Around each of these fixed points linear analysis is then possible for all $u$. The physical temperature $T$
introduces a cutoff scale for $u$ that allows us to find a coherent solution to the eigenvalue problem. Of course here $\epsilon=2$, hence the one-loop results are expected to
be approximate. Several recent works have found that the one-loop scheme provides
reasonable approximations of exponents and a clear, qualitatively correct picture
for this model \cite{alan,SchehrCO,Schehr05}.

\subsection{One-copy correlator at non-zero temperature $T$}
\label{sec_fix_temp}
 Temperature enters the FRG flow equation for the one-copy correlator in a natural way:
\begin{eqnarray}
\label{eq_flow_temp}
\partial_l \Delta_T(u) &=& \epsilon \Delta_T(u) - \Delta_T'(u)^2 \\ 
 && {} + \left[ T + \Delta_T(0) - \Delta_T(u) \right] \Delta_T''(u). \nonumber
\end{eqnarray}
The resulting fixed point equation is integrable:
\begin{eqnarray}
\label{eq_flow_T}
  \epsilon \Dfix_T(u) &=& \Dfix_T{}'(u)^2 \\ 
 && {} - (T + \Dfix_T(0) - \Dfix_T(u)) \Dfix_T{}''(u) \nonumber \\
&=& \frac 1 2 [(\Dfix_T(u) - \Dfix_T(0) - T)^2]'' \nonumber 
\end{eqnarray}
and is implicitly solved by quadrature \cite{ChauveBL}:
\begin{eqnarray}
\label{eq_fix_temp}
  u &=& \sqrt{\frac 3 {2\epsilon}} \; G(T,T + \Dfix_T(0) - \Dfix_T(u))
\end{eqnarray}
with
\begin{eqnarray}
  G(a,b) &\equiv& \int_a^b \frac{\dif y \; y}{\sqrt{(y-T)(y-y_-)(y_+-y)}}\\
  4y_{\pm} &=& 3\Dfix_T(0) + T \\
    && {} \pm \sqrt{3(3\Dfix_T(0)+T)(\Dfix_T(0)+3T)}\;, \nonumber \\
    && \quad \quad \text{with}\quad  y_- < 0 < T < y_+ \nonumber.
\end{eqnarray}
The constraint $\frac 1 2 = \sqrt{\frac 3 {2\epsilon}} G(T,y_+)$ yields the value of the fixed point correlator at zero $\Dfix_T(0)$ as a function of $T$.

The finite temperature correlator fixed point $\Dfix_T(u)$, given implicitly by equation (\ref{eq_fix_temp}) reduces to the non-analytic zero-temperature correlator of eq.~(\ref{eq_fix}) as $T \to 0$. Notice how the finite temperature $T$ rounds off the cusp within a boundary layer of width $\sim T$: The curvature at the origin becomes finite $\Dfix_T{}''(0) = - \frac {\epsilon\Dfix_T(0)} T$, and within the boundary layer, for $u\ll T$, the following series expansion holds:
\begin{eqnarray}
\label{eq_series_T}
  \Dfix_T(u) &=& \Dfix_T(0) - \frac{\Dfix_T(0) T} 2 \left(\frac u T\right)^2  \\
  & & {} \times \bigg[ 1 - \frac{3\Dfix_T(0) + T}{12}\left(\frac u T\right)^2 \nonumber \\
  & & {} \times \bigg[ 1 + \frac{15\Dfix_T(0) + T}{30}\left(\frac u T\right)^2 \bigg] \bigg] + O\left(\left(\frac u T\right)^8\right) \nonumber
\end{eqnarray}
These are the first terms of a systematic low temperature expansion \cite{ChauveBL}:
\begin{equation}
\label{eq_expand_fix} \Dfix_T(u) = \Dfix_T(0) - \sum_{k\ge1} T^k
\phi_k(u / T)
\end{equation}
with $\phi_k(0)=0$ valid inside the TBL, i.e. for $u/T = O(1)$. The
following matching conditions hold at large $x=u/T$:
\begin{eqnarray}
  & \phi_k(x) & \sim c_k |x|^k \\
  & \sum_{k\ge1} &  c_k |u|^k   = \Dfix_{T=0}(0) - \Dfix_{T=0}(u)
\end{eqnarray}
since for $u=O(1)$, $\Dfix_T(u) = \Dfix_{T=0}(u) +O(T)$. The first scaling form is
 $\phi_1(x)=\phi(x)-1$ where we define:
\begin{eqnarray}
\label{eq_scaling_first} \phi(x) = \sqrt{1 + \left(\frac{\epsilon
x}{6}\right)^2},
\end{eqnarray}
as can be verified by inserting the expansion of eq.~(\ref{eq_expand_fix}) into the FRG flow eq.~(\ref{eq_flow_temp}) and collecting orders in $O(T)$. The low temperature expansion will be
detailed and generalized in section \ref{all}, where $\phi_2$ will also be computed.
For the present purpose the following low temperature form of the
BL is sufficient:
\begin{eqnarray}
\label{eq_dfix_T}
 \lefteqn{ \Dfix_T(0) - \Dfix_T(u) = } \\
 && T \left( \sqrt{1 + \left(\frac{\epsilon u}{6T}\right)^2} - 1\right) + T^2 \phi_2(u/T) + O(T^3) \nonumber
\end{eqnarray}
It reproduces well the $T\rightarrow 0$ limit $\frac {\epsilon u} 6$, as well as the first term of the power series expansion around $u=0$ of eq.~(\ref{eq_series_T}).

\subsection{Linearization of non-zero $T$ flow equation}
\label{sec:linearT}

The FRG flow equation at $T>0$ for the two-copy correlator,
assuming the one-copy correlator has reached its fixed point
$\Dfix_T(u)$, reads:
\begin{eqnarray}
\label{eq_flow_TRP}
\partial_l D(u) &=& \epsilon D(u) - D'(u)^2 \\ 
 && + (T+\Dfix_T(0) - D(u))D''(u) \nonumber
\end{eqnarray}
with $\partial_l \Theta_l = - \partial_l D(0)$. Since
$D(u)=\Dfix_T(u)$ is now a (analytic) fixed point of this equation,
we define:
\begin{equation}
\label{deffT} D(u) = \Dfix(u) + f(u,l)
\end{equation}
and perform a linear analysis for small $f$, i.e. we write
$f(u,l)=e^{2 a l } f_a(u)$ and look for eigenfunctions. This can be
done at any $T$, at least numerically, using the implicit form of
the exact fixed point given in the last section. At low temperature
it can be done analytically, provided one distinguishes the two
regimes, $u \sim T$ (TBL) and $u \sim 1$. In the second regime the
analysis becomes identical, to leading order in $T$, to the one
performed directly at zero temperature in section \ref{sec:zeroT}.
In the TBL, it is natural to look for solutions of the form
$f_a(u)=\tilde f(u/T)$. The matching between the two regimes will be
studied in the next section. Inserting in the linearized
version of eq.~(\ref{eq_flow_TRP}) and (\ref{deffT}) and using the low $T$
expansion of eq.~(\ref{eq_expand_fix}) we obtain:
\begin{eqnarray}
  0  &=& (\epsilon-2a) \tilde f(x) \\ 
 && {} + \frac {\epsilon^2}T \frac{\dif^2}{\dif x^2}[(\phi(x)+T\phi_2(x)) \tilde f(x)] \nonumber
\end{eqnarray}
from which a systematic low $T$ expansion of $\tilde f(x)$ can be
obtained.

For $a=\epsilon/3$, one finds that a simple ansatz almost solves it
(in an approximate sense given below) namely:
\begin{equation}
\label{eq_ansatz} \tilde f(x) = \frac{1}{\phi(x)} -\frac{12
T}{\epsilon} \ln(\phi(x)).
\end{equation}
noting that, from eq.~(\ref{eq_scaling_first}):
\begin{subequations}
\begin{eqnarray}
  \phi'(x) &=& \frac x {36 \phi} \; = \; \frac{\sqrt{\phi^2 -1}}{6\phi} \\
  \phi''(x) &=& \frac 1 {36\phi}\left( 1 - \frac{x\phi'}{\phi}\right) \; = \; \frac 1 {36\phi^3}
\end{eqnarray}
\end{subequations}
inserting into the right hand side of the above equation leads to:
\begin{eqnarray}
  0 &=& (1-2/3)(1/ \phi - \frac{12 T}{\epsilon} \ln \phi) \\
   && {} + \frac {\epsilon} T [1-\frac{12T}{\epsilon}\phi \ln \phi 
         + \frac{\phi_2}{\phi}]''  \nonumber\\
  &=& \frac 1 {3\phi} - 12\left[ \phi''(1+\ln \phi) + \frac{\phi'^2}{\phi}\right] + \epsilon\left(\frac{\phi_2}{\phi}\right)'' + O(T)  \nonumber\\
 &=& \frac 1 {3\phi} - \frac 1 {3\phi} + \frac{\ln \phi}{3 \phi^3} + \epsilon\left(\frac{\phi_2}{\phi}\right)'' + O(T) \nonumber
\end{eqnarray}
Hence the dominant terms cancel, and the remaining term $\frac{\ln
\phi}{3 \phi^3}$ is found to be subdominant at large $u$ going as
$\sim \frac{\ln u}{u^3}$ as $u\rightarrow\infty$. A more complete
analysis is done in section \ref{sec_tbl_one_copy}, where it is shown that the term
$\epsilon(\phi_2/\phi)''$ is also subdominant.
For now suffice it to note that the form of eq.~(\ref{eq_ansatz}) is exact
to dominant order in $T$ for all $x$, and to next order in $T$
it reproduces the exact large $x$ behavior. Thus, surprisingly, we
find evidence for a {\em logarithmic} correction to the
eigenfunction, emerging at non-zero temperature $T>0$.

\subsection{Logarithmic temperature dependence of the eigenvalue}
\label{subsec:eigen} 

Let us examine the correction, induced by temperature, to 
the eigenvalue. Assume that the
main contribution to the eigenvalue correction comes, to first
order, from the much larger regime outside the boundary layer where
the eigenfunction is given by the expression of eq.~(\ref{eq_evector}). Assume 
further that $a$ is close to the value $a=\epsilon/3$ and expand the
expression for the corresponding eigenfunction in powers of $\delta
a = \frac a {\epsilon} - \frac 1 3$, as well as around $u=0$, using
eq.~(\ref{expansion}):
\begin{eqnarray}
\label{eq_f_expansion}
  f_a(u) &=&  1  + \delta a \left( \frac 1 u - 3 - 2 \ln(4u) \right) \\ 
  && {} + O(\delta a^2,u^2,\delta a \,u)  \quad , \quad u=O(1) \nonumber
\end{eqnarray}
The divergence at $u=0$ is rounded off inside the TBL.
The previous paragraph gives the expression
inside the TBL, to leading order in $T$:
\begin{eqnarray}
f_a(u) = \frac{\delta a\; \epsilon}{6T} \frac{1}{\sqrt{1 +
\left(\frac{\epsilon u}{6T}\right)^2}}  \quad , \quad u=O(T)
\end{eqnarray}
where we have multiplied with a constant in order to match
 the $1/u$ term of eq.~(\ref{eq_f_expansion}) for large $u/T$.

The eigenfunction is now integrable and we enforce the
condition of zero mean, expressed as $\int_0^{1/2} \dif
u f_a(u) = 0$ when taking into account that the eigenfunction is symmetric about $u=1/2$. We
split the integral into two parts, inside $u<k T$ and outside $u> k
T$ the boundary layer (any large constant $k$ will do), and we
find up to order $O(T,\delta a)$:
\begin{eqnarray}
  0 &=& \int_0^{\frac 1 2} \dif u f_a(u)  \\
    &=& \frac{\delta a\; \epsilon}{6T}\int_0^{k T} \frac{\dif u}{\sqrt{1 + \left(\frac{\epsilon u}{6T}\right)^2}} \nonumber \\
    && {} + \int_{k T}^{\frac 1 2} \dif u \left( 1 + \delta a \left( \frac 1 u - 3 - 2 \ln 4u + O(u) \right) \right) \nonumber \\
\label{eq_int_zero}
      0 &=& \hspace{1.75cm} \frac 1 2 - \delta a \;(\ln T + O(1)) + O(T). \nonumber
\end{eqnarray}
This shows that for the eigenfunction to integrate to zero, i.e. for
the $\ln T$ divergence to be compensated, the eigenvalue acquires a
logarithmic temperature dependence.
\begin{equation}
  \label{eq_a_T}
  \delta a = - \frac 1 {2 \ln \frac 1 T} + O(\delta a^2)
\end{equation}
In fact, inserting this expression for the eigenvalue correction
back into the eigenfunction expansion of eq.~(\ref{eq_f_expansion}) and
normalizing by $6T/\delta a \epsilon$, we retrieve exactly the
asymptotic form of the eigenfunction inside the boundary layer (eq.~(\ref{eq_ansatz})).
\begin{eqnarray}
\label{eq_eigen_anal}
 & & \frac{6T}{\delta a \epsilon} f_a(u) = \frac{6T}{\epsilon u} - \frac{12T}{\epsilon} \; \ln \frac u T + \ldots\\
  && {} = \lim_{\frac u T \rightarrow \infty} \left( \frac 1 {\sqrt{1+\left(\frac{\epsilon u}{6T}\right)^2}} - \frac{12 T}{\epsilon} \; \ln \sqrt{1+\left(\frac{\epsilon u}{6T}\right)^2} \right) \nonumber
\end{eqnarray}
including all logarithmic terms. Hence our solution satisfies the
required conditions, zero mean and matching between inside and
outside the TBL. Note that it was necessary to not only include the dominant
contribution in the TBL, but also the subdominant one to logarithmic
accuracy.

In principle the low T expansion can be pursued to higher orders.
The second order $O(\delta a^2)$ is much harder to calculate though,
because it requires us to exactly integrate the non-divergent part
of $f_a(u)$ over the whole interval $[T,1/2]$.

\section{numerical study of the random periodic class at $T>0$ ($\theta=0$,
$d=2$)}  \label{sec:num} 

To test the subtle mechanism for the selection of the
eigenvalue based on the boundary layer matching, we now turn to a
numerical analysis.

\subsection{Shooting to solve eigenvalue problem}

We numerically solve the FRG flow equation for the two-copy
correlator (eq.~(\ref{eq_flow_TRP})) linearized around the exact
implicit solution of the one-copy correlator at finite $T$ (eq.~(\ref{eq_fix_temp})). The eigenvalue equation is
\begin{eqnarray}
  \label{eq_lin_T_exact}
  0 &=& (\epsilon - 2 a - \Dfix_T{}''(u))f_a(u) - 2 \Dfix_T{}'(u) f_a'(u) \nonumber \\ 
 && {} + (T + \Dfix_T(0)- \Dfix_T(u)) f_a''(u) 
\end{eqnarray}
with periodic boundary conditions: $f_a(0) = f_a(1) = 1$ and
$f_a'(0) = f_a'(1) = 0$. Again, since the eigenfunction is even
about $u=1/2$, it suffices to consider the interval $u \in [0,1/2]$,
requiring $f_a'(1/2) = 0$. We set $\epsilon$ to unity since it
plays no role.

\subsubsection{Numerical details}

We need a continuous numerical representation for the fixed point
correlator $\Dfix_T(u)$, whose analytical properties are given in section \ref{sec_fix_temp}. 
The first step is to evaluate $\Dfix_T(0)$ for
given temperature $T$, from the constraint $1/2 = \sqrt{3/2}\;
G(T,y_+)$. We have to find $\Dfix_T(0)$ such that the elliptic
integral is exactly one half. This is easily done using the Brent
method. The integrand has a divergence $\sim 1/\sqrt{u}$ at each
limit, which is handled by a change of variables to $v^2 = u$. We
use a Romberg integration routine to efficiently get the desired
precision.

We then use the implicit expression of eq.~(\ref{eq_fix_temp}) to
calculate a discrete representation $\{u_i(\Delta_i),\Delta_i; i =
1..N\}$ of the correlator. The $\Delta_i$ are chosen such that we
have a sufficiently fine discretization inside the boundary layer $u
\sim T$, even for very small values of $T$, and are chosen less
dense outside the boundary layer, to reduce the total number $N$ of
support points.

A cubic spline interpolation of the discrete function
$\Delta_i(u_i)$ allows us to obtain a continuous representation of
the correlator $\Dfix_T(u)$. At the origin $u=0$ we take advantage
of the exact series expansion of the correlator of eq.~(\ref{eq_series_T}), matching the series expansion to the spline at
about $u\sim0.01T$.

With this continuous numerical representation of the correlator
fixed point, we can solve the eigenvalue problem of eq.~(\ref{eq_lin_T_exact}) 
to arbitrary precision, using the shooting
method: make an initial guess for the eigenvalue $a$, and integrate
the ordinary differential equation starting from the initial
condition $f_a(0) = 1, f_a'(0) = 0$, by means of a standard
integration routine (e.g. odeint). Aiming for periodic boundary
conditions $f_a'(1/2) = 0$, one finds the eigenvalue $a(T)$.

In this way, we calculate the eigenvalues $a(T)$ and
eigenfunctions $f_a(u)$ as a function of the temperature $T$ to
arbitrary precision. We need however to go to very small values of
$T\simeq 10^{-16}$, to clearly see the logarithmic dependence on $T$
of the eigenvalue (eq.~(\ref{eq_a_T})). This requires quadruple
precision for the numerics.

\subsection{Numerical results}

\begin{center}
  \begin{figure}
    \centerline{\includegraphics[bb=86 75 337 251]{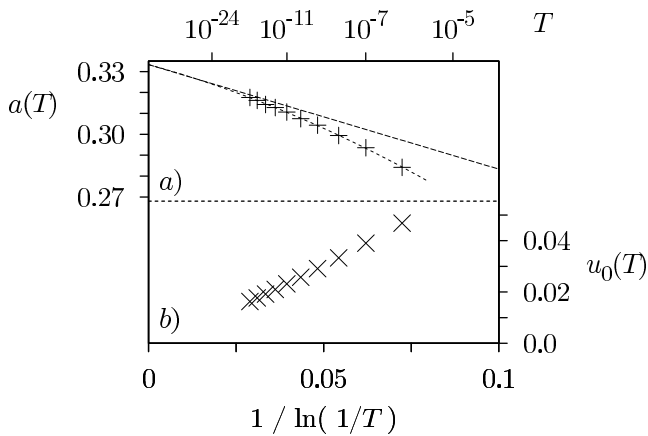}}
    \caption{{\em Slow logarithmic dependence on cutoff $T$.} $a)$($+$) Eigenvalue $a$ and $b$)($\times$)
    zero $u_0$ of the first eigenfunction approach
    their zero temperature values $a)1/3\; b)0$ as $1/\ln[1/T]$ (we set $\epsilon=1$).
    The linear coefficient of the fitted numerical
    eigenvalue data coincides with the theoretical value
    of $-1/2$. Equally, the linear coefficient of the fitted zero
    of the eigenfunction is identical to the theoretical value of $1/2$.
    The ordinate scale is linear in $1/\ln[1/T]$. As a guide, the corresponding
    value for $T$ is given on the upper ordinate scale.}
    \label{fig_a_T}
  \end{figure}
\end{center}

Indeed, we find numerically that the largest physical eigenvalue
$\frac a {\epsilon}$ is equal to one third plus corrections
logarithmic in the cutoff $T$ (see figure \ref{fig_a_T}, upper
half). This confirms our analytical finding of logarithmic
corrections to the eigenvalue. There exists one larger eigenvalue
$\frac a {\epsilon} = 1/2$. It is easy to see that the exact eigenfunction for
this eigenvalue is:
\begin{eqnarray}
&& f_{a=\epsilon/2}(u) = \frac{K}{\Delta_T^*(0) - \Delta_T^*(u) + T}
\end{eqnarray}
for all $u$ inside and outside the TBL (which correctly matches 
the eigenfunction $f_{a=\epsilon/2} \sim 1/(u(1-u))$ for $u \gg T$). This
eigenvalue does not acquire any corrections in
the cutoff $T$, and most importantly, its corresponding
eigenfunctions are strictly positive, they do not have any zeros.
This means they cannot have zero mean and hence do not correspond to
a correlator of the RP class.

\begin{center}
  \begin{figure}
    \centerline{\input{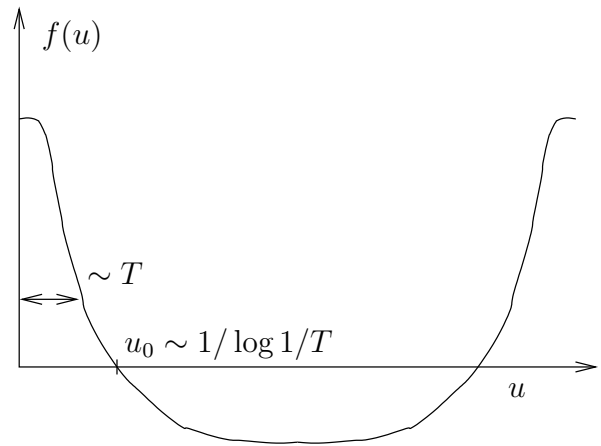}}
    \caption{{\em Different regimes of eigenfunction.} The eigenfunction has three different regimes, separated by two crossover values, represented schematically here. Near zero, the eigenfunction presents a boundary layer
      of width $u\sim T$ as found in Fig. \ref{fig_efunction}. Outside the boundary layer, the eigenfunction falls off, and crosses zero at $u_0$. This zero has a logarithmic dependence on temperature, introducing a second crossover, which scales differently from the width of the boundary layer $u_0 \sim 1/\log 1/T$. For even larger values of $u$, the eigenfunction flattens but remains negative, such that the overall integral of the RP eigenfunction is strictly zero.} \label{fig_schema}
  \end{figure}
\end{center}

\begin{center}
  \begin{figure}
    \centerline{\includegraphics[bb=86 75 337 251]{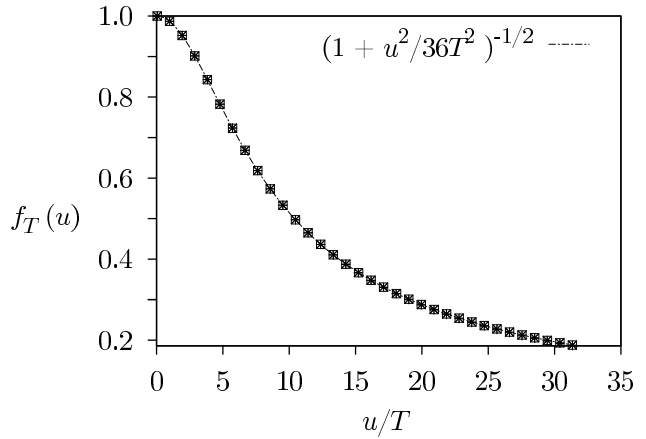}}
    \caption{{\em Scaling form of eigenfunctions.} Inside a boundary layer
    of width $u\sim T$, the eigenfunctions present a functional
    form independent of cutoff $T$. Numerical data for various
    temperatures $T=10^{-4}...10^{-8}$ is indistinguishable
    from the theoretical expression $f_T(u) = 1/\sqrt{1 + u^2/36T^2 }$.}
    \label{fig_efunction}  
  \end{figure}
\end{center}

We recall the first analytical terms of the eigenfunction of eq.~(\ref{eq_eigen_anal}, with $\epsilon=1$):
\begin{eqnarray}
  \label{eq_evector_limit}
  && f_T(u) = \frac 1 {\sqrt{1+\frac{u^2}{36T^2}}} - 12 T \; \ln \sqrt{1+\frac{u^2}{36T^2}} \\
 &&  \lim_{u/T \rightarrow \infty} f_T(u) = \frac{6T}u - 12 T \; \ln \frac u T \nonumber
\end{eqnarray}
This expression compares rather well to the numerically calculated
solution (figure \ref{fig_efunction}). 
Moreover, setting equation (\ref{eq_evector_limit}) to zero gives us the first order
term of the zero of the eigenfunction $u_0 = \frac 1 {2 \ln 1/T}$.
The prefactor $\frac 1 2$ is exactly the one found in the numerical
data of the first zero of the eigenfunction (see figure \ref{fig_a_T}, lower
half). Thus the numerical
results confirm our analytical analysis, providing us with a
coherent picture of the solution of the linearized FRG flow equation
at non-zero $T$. This is represented schematically in Fig. \ref{fig_schema}.

As the cutoff $T$ approaches zero, the
eigenfunction shifts more and more weight into the ever smaller BL,
in order to still fulfill the zero-mean constraint $\int f = 0$.
While this picture is satisfactory for any fixed $T$ it is still not
clear whether it could help to solve the problem directly at $T=0$.
The question of how in the limit $T \rightarrow 0$ this
eigenfunction develops the non-analyticity necessary to balance the
cusp in the one-copy correlator is especially subtle given that
there are two regimes with different $T$-scaling properties. The
boundary layer disappears as $u_{\msmall \text{BL}} \sim  T$,
whereas the zero of the eigenvector approaches zero as $u_0 \sim
1/\ln[1/T]$ (figure \ref{fig_a_T}, lower half). 

\subsection{Consequence of logarithmic correction}

The unusual logarithmic correction to the eigenvalue --- caused by a
finite cutoff length --- implies that it is very hard to calculate
the latter by means of intuitive numerical approaches. If for
example one simply tries to numerically integrate the FRG flow
equation, one necessarily introduces a cutoff length $(\sim N^{-1},
\; N$ being the number of discretization intervals in real space, or
the highest frequency mode in Fourier space). This cutoff length has
exactly the same effect as the finite temperature cutoff, i.e.
preventing access to smaller lengthscales. Thus even at zero real
temperature, any finite numerical cutoff introduces an immediate and
non-negligible correction to the eigenvalue of order $\ln[N]^{-1}$.
For example, a reasonably large $N \approx 10^6$ leads to a
correction of order $10\%$ to the eigenvalue.

A logarithmic correction to an eigenvalue,
not unsimilar to the present situation,
has been noticed in the context
of a propagating wave front \cite{DerridaBrunet}.

\section{General analysis at $T>0$ and extension to other classes}
\label{all}

We turn from the simple case ($d=2$, RP class) where the
temperature does not flow, to the general problem at non-zero 
temperature. The two equations (\ref{eq_delta_flow})
are studied with 
the temperature allowed to flow, i.e. $\theta>0$.
Despite $T_l=T e^{- \theta l}$ flowing to zero, it
is still possible to use linear analysis, as we will show. We 
look again for a zero mean eigenfunction, placing us in the RP 
class for $d>2$, and the RB class.

\subsection{TBL for one copy correlator}
\label{sec_tbl_one_copy}
We start by solving more accurately the first equation 
(\ref{eq_delta_flow}) and write:
\begin{subequations}
\begin{eqnarray} \label{ansatztbl} 
 \Delta(u) &=& \Delta^*(u) \quad , \quad u = O(1) \\
 \Delta(u) &=& \Delta(0) - T_l \phi_1(\chi u/T_l) - T_l^2
  \phi_2(\chi u/T_l) \\ 
 && + O(T_l^3) \quad , \quad u = O(T_l) \\
 \Delta(0) &=& \Delta^*(0) - T_l \gamma_1 + O(T_l^2)
\end{eqnarray}
\end{subequations}
with $T_l = T e^{-\theta l}$. One must have $\phi_n(0)=0$, and 
$\phi_1(x) \sim |x|/6$ at large $x$ to fit the cusp, hence
the choice $\chi=6 |\Delta^{* \prime}(0^+)|$. 
The zero temperature
FP $\Delta^*(u)$ can now be of RP type, as well as RB (in which case
$\zeta$ is non-zero and determined by the FP equation and the SR
boundary condition). From the analysis
performed in Appendix \ref{app:tbl} one finds:
\begin{eqnarray}
&& \phi_1(x) = \sqrt{1+x^2/36} - 1 = \phi(x)-1
\end{eqnarray}
and $\chi^2 = \epsilon^2 \tilde \chi^2 = 36 \Delta'(0^+)^2 = 36
(\epsilon-2\zeta) \Delta^*(0)$ ($\tilde \chi=1$ for the periodic FP)
using $\phi_1''(0)=1/36$. From which we recover the zero temperature
fixed point (eq.\ref{eq_fix}) in the large argument limit 
$[\phi_1(x\rightarrow \infty) = \frac{|x|}6 + \frac 3{|x|} + \ldots
\;]$. To next order one finds:
\begin{eqnarray}
 \phi_2(x) &=& \frac 1 {\chi^2 \phi(x)} [ 12 (6 - 2 \epsilon + 2
\zeta) (\phi(x)-1) \\
 && {} + x^2 (1-\gamma_1) + \frac{1}{6}
  x^2 \phi(x) (\zeta - \epsilon)) \nonumber \\ 
 && {} - 3(4 + \zeta - \epsilon)\,x\,\text{ArcSinh}(\frac{x}{6}) ) ] \nonumber
\end{eqnarray}

Note that $\phi_2(x) \sim (\zeta - \epsilon) x^2/(6 \chi^2)+ O(x)$
at large $x$. This is compatible with $\Delta^{* \prime
\prime}(0^+)=(\epsilon-\zeta)/3$. Note also that:
\begin{eqnarray}
\lim_{x\rightarrow \infty} \left(\frac{\phi_2(x)}{\phi(x)}\right)''
&=& \frac{72}{\chi^2 x^3}\bigg( 15\theta + 7\epsilon - 22 \zeta \\ 
&& {} + 3 (\epsilon + 2 \theta - 3 \zeta) \ln (\frac{3}{x}) \bigg) + O(x^{-4}) \nonumber
\end{eqnarray}
as promised in section \ref{sec:linearT}, hence validating the
approximate solution given there.

\subsection{Equation for the two copy correlator}

Now we define the solution of the second equation in
(\ref{eq_delta_flow}) to be:
\begin{eqnarray}
&& D(u) = \Delta(u) + f(u)
\end{eqnarray}
and we study the resulting equation for $f(u)$ in a linear
approximation:
\begin{eqnarray}
  \label{eq_lin_T}
  \partial_l f(u) &=& (\epsilon - 2 \zeta) f(u) +  \zeta u f'(u) \\
  && {} +  \frac{d^2}{d u^2} \big[ \big( \;\; T_l + \Delta(0)- \Delta(u)\;\; \big) f(u) \big] \nonumber
\end{eqnarray}
The only neglected term is $- \frac{1}{2} \frac{d^2}{d u^2} f(u)^2$
on the r.h.s. One can write:
\begin{eqnarray}
  \partial_l f(u) =& (\epsilon - 2 \zeta) & f(u) +  \zeta u f'(u) \\
 & {} + \frac{d^2}{d u^2} \big[\big( & T_l \phi(\chi u/T_l) + T_l^2
\phi_2( \chi u/T_l) \nonumber \\ 
 && {} + O(T_l^3) \;\; \big) f(u) \big]  \quad , \quad u \sim T_l \label{in} \nonumber \\
  \partial_l f(u) =& (\epsilon - 2 \zeta) & f(u) +  \zeta u f'(u) \\ 
 & {} + \frac{d^2}{d u^2}
  \big[\big( & \Delta^*(0)- \Delta^*(u) \;\; \big) f(u) \big] \nonumber \\ 
 & {} + O(T_l)& \hspace{3cm} , \quad u \sim O(1) \label{out} \nonumber
\end{eqnarray}
where we have explicitly separated the inside from the outside of the
TBL.

\subsubsection{Eigenfunction inside the TBL}

Inside the TBL we look for a solution of the form:
\begin{equation}
 f(u) = \frac{\chi}{6 T_l} \tilde f(x= \chi u/T_l) \quad , \quad u \sim T_l
 \label{ansatzTL}
\end{equation}
As we will see below, the prefactor $1/T_l$ is crucial for obtaining a
correct matching to the $u=O(1)$ regime.

Equation (\ref{in}) gives:
\begin{eqnarray}
\partial_l \tilde f(x) &=& (\epsilon - 2 \zeta - \theta) \tilde f(x) + (\zeta-\theta) x
\tilde f'(x) \\ 
 && {} + \frac{\chi^2 }{\tilde T_l} \frac{d^2}{d x^2}
  \Big[ \Big( \;\; \phi(x) + T_l
\phi_2(x) + O(T_l^2) \;\; \Big) \tilde f(x) \Big] \nonumber
\end{eqnarray}
The solution seems to admit the expansion:
\begin{eqnarray}
&& \tilde f(x) = e^{2 a l}  \left( \frac{1}{\phi(x)} + T_l \psi(x) +
O(T_l^2) \right)
\end{eqnarray}
where:
\begin{eqnarray}
 0 &=& (\epsilon - 2 \zeta - 2 a - \theta) \frac{1}{\phi(x)} -
(\zeta - \theta) x \frac{\phi'(x)}{\phi(x)^2} \nonumber \\ 
 && {} + \chi^2 \frac{d^2}{dx^2} \left[ \frac{\phi_2(x)}{\phi(x)} + \phi(x) \psi(x)  \right]
\end{eqnarray}
This yields:
\begin{eqnarray}
 \psi(x) &=& \frac{\alpha}{\phi(x)} - \frac{\phi_2(x)}{\phi(x)^2} +
\frac{1}{\chi^2} \bigg(36 ( \epsilon + \theta - 4 \zeta - 2 a) \nonumber \\
 && {} + (2 a - \epsilon  + 3 \zeta) \frac{6 x \text{ArcSinh}(\frac{x}{6})}{\phi(x)}
\bigg)
\end{eqnarray}
with $\alpha$ undetermined. At large $x$ one has:
\begin{eqnarray}
  \psi(x) &=& \frac{36}{\chi^2} (2 a - \epsilon  + 3 \zeta)
\ln\left(\frac{x}{3}\right) \\ 
 && {} +  \frac{36}{\chi^2} ( \epsilon + \theta - 4 \zeta
- 2 a) - \frac{6}{\chi^2} (\zeta-\epsilon) + O(1/x) \nonumber
\end{eqnarray}
If one defines the ''eigenfunction'' $f_a(u)$ through $f(u)=e^{2 a l} f_a(u)$,
it has the following behavior at large $u/T_l$:
\begin{eqnarray}
f_a(u) \sim \frac{\chi}{6 T_l} \left[ \frac{6 T_l}{\chi u} +
\frac{36}{\chi^2} (2 a - \epsilon  + 3 \zeta) T_l \ln\left(\frac{u}{3
T_l}\right) \right] \label{large}
\end{eqnarray}
consistent, up to a normalization with the result (\ref{eq_eigen_anal}) for the RP class in $d=2$, setting $\zeta=0$, $\chi=\epsilon$, $a=1/3$. This expression will be
matched to the small $u$ behavior of the
eigenfunction $f_a(u)$ in the regime $u=O(1)$, which contains a small $1/u$ term.
Note however that $f_a(u) \sim (\chi/(6 T_l))/\sqrt{1+\chi^2 u^2/(36 T_l^2)}$ in the TBL,
hence $f_a(0)$ has a different dependence in $l$ that $f_a(u)$ for $u=O(1)$, i.e.
it is a non-uniform eigenvector. 

\subsubsection{eigenfunction outside the TBL}

For $u$ of order unity one must study the linear differential
equation (\ref{out}) containing the zero temperature fixed point
$\Delta^*(u)$. 

Let us start with the RP class (for $d>2$). Since the fixed point $\Delta^*(u)$ does not
change form (apart from the overall factor of $\epsilon$ already taken
into account) we expect the same behavior as for $d=2$. From
the ansatz of eq.~(\ref{ansatzTL}) one sees that for $\zeta=0$ the behavior
(eq.~(\ref{large})) identifies with eq.~(\ref{eq_eigen_anal}) which matches
the one for $u=O(1)$, (eq.~(\ref{eq_f_expansion})), as discussed already in
section \ref{subsec:eigen}. Hence for the RP class 
\footnote{Note again for $\zeta=0$ the exact eigenfunction of (\ref{equ})
$f_{a=\epsilon/2}(u) = \frac{K}{\Delta^*(0) - \Delta^*(u)}$
which does not satisfy the required conditions.},
we find the same growth exponent $a=
\epsilon/3$ for the function outside the TBL (for $u=O(1)$). But from the
discussion of the end of the previous paragraph, the growth exponent of:
\begin{eqnarray}
&& \Theta_l \sim e^{2 \tilde a l}
\end{eqnarray}
is determined by $f(0)$ inside the
TBL, hence $\tilde a = a + \frac{\theta}{2}$. This non-uniformity is
the main difference with the case $d=2$.

For the RB class the equation for the eigenvector for $u=O(1)$ reads:
\begin{eqnarray} \label{equ} 
 0 &=& (\epsilon - 2 \zeta - 2 a) f(u) +  \zeta u f'(u) \\
 && {} +
 \frac{d^2}{d u^2}
  \big[\big( \;\; \Delta^*(0)- \Delta^*(u)\;\; \big) f(u) \big] \nonumber
\end{eqnarray}
This equation involves the RB fixed point, which is non trivial
\cite{fisher} and was determined with high accuracy in
\cite{twoloopdep} together with the value for $\zeta=0.208298063$. 
An analysis near $u=0$ shows that any solution of eq.~(\ref{equ}) 
has the form of eq.~(\ref{expansion}) at small $u$, with 
$B_a=(\epsilon - 3 \zeta - 2 a)/\Delta^{* \prime}(0^+)$. 
We will assume that one can then proceed as for the RP case 
and look for the value $a=a_{\text{RB}}$ such that $A_{a_{\text{RB}}}=0$.
This is equivalent to the shooting problem of solving
eq.~(\ref{equ}) imposing that $f(0)=1$ and that $f(u)$ decay at infinity.
This fixes a unique and non-trivial value for $a_{\text{RB}}$. 
We have solved this shooting problem numerically using Mathematica \footnote{We thank Kay Wiese for
providing high precision approximations for $\Delta^*$.},
and found $a_{\text{RB}} = 0.083346(6) \epsilon$. The corresponding 
eigenvector is everywhere positive and integration of eq.~(\ref{equ})
easily leads to the following constraint:
\begin{eqnarray} \label{equ2} 
&& (\epsilon - 3 \zeta - 2 a) \int_{0^+}^\infty du f(u) = - \Delta'(0^+) f(0).
\end{eqnarray}
Indeed one finds $a_{\text{RB}} < (\epsilon - 3 \zeta)/2$ (the $0^+$ means that
the domain excludes the TBL).
Once this eigenfunction is determined in the region $u=O(1)$
the method to satisfy the zero integral condition over the full axis including
the TBL is the same
as for the RP class. First, one checks that the solution
(eq.~(\ref{large})) in the TBL matches correctly the 
small $u$ behavior (eq.~(\ref{expansion})) of the $u=O(1)$ regime, using
$B_a=(\epsilon - 3 \zeta - 2 a)/\Delta^{* \prime}(0^+)$.
Second, one has again $A_{a_{\text{RB}} + \delta a} \sim \delta a$
and one can proceed as in section \ref{subsec:eigen}.
One finds that the zero mean condition again leads to logarithmic
corrections $a_l = a_{\text{RB}} - K/ln(1/T_l) = a_{\text{RB}} - K/(\theta l)$
where $K = \int_{0^+}^\infty du f(u)/\partial_a A_a|_{a=a_{\text{RB}}}$ given by 
eq.~(\ref{equ2}). The main difference with the RP class, besides the value of
the growth exponent, is that the logarithmic temperature corrections are also 
(weakly) $l$-dependent \footnote{This dependence
being much weaker than the leading exponential one, it implies an additional
$l$ dependence of the eigenvector which is subdominant and neglected here.}. 

\section{Summary and Discussion}

In this paper we have studied the problem of two mani-folds pinned in slightly 
different random potentials at the same temperature. We have written the coupled FRG equations for
the single and two-copy correlators, $\Delta(u)$ and $D(u)$, and temperature, 
to one loop accuracy.
We have investigated how the difference between the two copies increase 
with scale. We have focused on zero-momentum ($q=0$) quantities and specifically we have
computed the fluctuations of the difference in (thermally averaged)
center-of-mass positions $\bar u^i(v)$, $i=1,2$, of the two copies, in the presence of a uniform confining
harmonic potential (of curvature $m^2=m_0^2 e^{-2 l}$) centered at a common position $u=v$. 
This observable is exactly given by:
\begin{eqnarray} \label{deltadef2}
&\frac{1}{2}& \overline{ (\bar u^1(v) - \bar u^2(0) - v)^2 } \\
 &&= A_d (L m)^{-d} m^{-2 \zeta }(\Delta(v m^{\zeta} ) - D(v m^{\zeta})) \nonumber
\end{eqnarray}
and can be seen to measure the r.m.s shift in position of $(L/L_m)^d$ roughly independent pieces
of manifolds of typical size $L_m \sim 1/m$ (hence the factor $(L m)^{-d}$ from the
central limit theorem). The deviation of the one-copy center of mass from the center of the well
is typically $\bar u^i(v) - v \sim O(m^{-\zeta})$. We have defined the two growth exponents:
\begin{eqnarray} \label{deltadef3}
 \Theta_l := \Delta(0) - D(0) &=& \tilde C e^{2 \tilde a l} \\
 \Delta(u) - D(u)|_{u = O(1)} &=& C e^{2 a l} f_a(u) \label{87}
\end{eqnarray}
allowing for the possibility that $\tilde a \neq a$. The coefficients $C$ and $\tilde C$
vanish when the difference in disorder $\delta$ between the two copies is taken to zero. 

We have obtained these exponents at non-zero temperature, using 
linear analysis and matching the regime $u=O(1)$ to the regime $u \sim T_l=T e^{-\theta l}$
in eq.~(\ref{87}), corresponding respectively to $v=O(m^{-\zeta})$ and 
$v = O(T m^{\theta-\zeta})$, the width of shocks, in eq.~(\ref{deltadef2}). For the random periodic class for
$d \geq 2$ and for the random bond class we found, with $\epsilon=4-d$:
\begin{eqnarray} \label{growth_exponents}
\tilde a &=& a + \frac{\theta}{2} \\
 a_{\text{RP}} &=& \left(\frac{1}{3} - \frac{1}{2 \ln(1/T_l)}\right) \epsilon \\
 a_{\text{RB}} &=& \left(0.083346(6) - \frac{K}{\ln(1/T_l)}\right) \epsilon
\end{eqnarray}
The result for the $d=2$, RP class has been thoroughly checked via a
numerical simulation. The other results assume that the scenario 
demonstrated in that case can be extended, which appears to be 
consistent. To confirm it further would require extensive numerics
or a more complete analytical study.

Since it originates from linear analysis, the growth of eq.~(\ref{deltadef3}) is valid only
for a limited range of scales $l=\ln(m/m_0)$ (i.e. of masses $m$ in eq.~(\ref{deltadef2})). 
From eq.~(\ref{eq_lin_T}), where the
term $(f^2)''$ has been neglected the linear analysis is valid only, in the regime $u=O(1)$, as long as
$C e^{2 a l} \ll 1$. Beyond that scale the growth is non linear. 
However, since the eigenvector is peaked in the TBL 
the condition of validity of our analysis is more stringent. One can write the condition
$(f^2)'' \ll T_l f''$ and substitute $f \sim (C/u) e^{2 a l}$ with $u \sim T_l$. 
This yields $C e^{2 a l}/T_l \sim \tilde C e^{2 \tilde a l}  \ll T_l$ which, not surprisingly, can be 
written as:
\begin{eqnarray} \label{thermal}
&& \Theta_l \ll T_l
\end{eqnarray}
i.e. the width of the chaos BL (which exists at $T=0$) is smaller
than the TBL. This is the condition for validity of the thermal regime
and linear analysis. Qualitatively, it means that the thermal width of the shocks 
should be larger than the typical shift in their relative position in the two copies.

More issues remain to be understood. If one defines the overlap length 
$L_\delta \sim e^{l_\delta}$ from $\Theta_{l_\delta} \sim 1$ 
one can argue that the chaos exponent is $\alpha=\tilde a$ in the thermal regime. Note however that
this assumes that $\tilde C \sim \delta^2$, a natural condition,
but which may be spoiled at scales around the
Larkin length where we do not have good control on the flow. 
Concerning the definition of $L_\delta$ from eq.~(\ref{chaoscorr}), 
using eq.~(\ref{general}) as was discussed there, we need a
more precise calculation of the scale dependence of the
non-local terms in the FRG. Finally,
the present analysis does not solve the
question of the zero temperature chaos boundary layer
when the condition of eq.~(\ref{thermal}) is violated,
although it gives some insight into selection mechanisms
for growth eigenvalues in such FRG equations. 

To conclude, we have made a step towards 
solving the intricate non-linear coupled FRG flow of the force
correlators. We have found an interesting
result for the growth exponent of the elastic manifold in frozen disorder
when the problem can be solved using a linearized flow equation. 
We have shown that it acquires a surprisingly
large logarithmic correction when a finite cutoff length is present
(here the temperature). More work is necessary to completely solve the problem and
to establish the exact relation between the eigenvalue found here and the
chaos exponent.

\appendix

\section{analysis of linearized equation}
\label{appendix1}

We give a detailed account of the linearized flow equation at zero 
temperature (eq.~(\ref{eq_flow_lin})).
As we are looking for a continuous solution periodic on $u \in [0,1]$, we are only interested in the even solutions. The first solution that springs to mind is $f(u) = 1/u(1-u)$, corresponding to the eigenvalue $a = \epsilon/2$. It is however not integrable at $u=0$.

In order to get an idea of possible integrable eigenfunctions, we look for finite series solutions. Applying the Frobenius method, we postulate a power-series solution of the form
\begin{equation}
  f_a(u) = u^r \sum_{n=0}^{\infty} c_n u^n
\end{equation}
insert it into equation (\ref{eq_flow_lin}) and equate coefficients of each term in the power series. The lowest power (indicial equation) provides the two possible values for $r=0,-1$.

In the case of $r=0$, the recurrence relation for the coefficients is:
\begin{equation}
c_{m+1} = c_m \frac{m(m+3) - 4 -12a/\epsilon}{(m+1)(m+2)}
\end{equation}
Notice that the ratio of coefficients tends to one $\lim_{m\rightarrow\infty} \frac{c_{m+1}}{c_m} = 1$, hence the infinite series does not converge uniformly on the interval $[0..1]$.
We thus have to demand that the series terminate after a finite number of terms, i.e. that the $m^{th}$ coefficient be zero $c_{m+1}=0$, giving us the eigenvalues $\tilde a_m$:
\begin{equation}
\label{eq_evalues}
  \tilde a_m = \frac{\epsilon} 3 \left( 1 - \frac{m(m+3)} 4 \right)
\end{equation}
Here we include even and odd solutions, whereas the eigenvalues $a_n$ of section \ref{sec:zeroT}
 are $a_n = \tilde a_{m=2n}$.
The case $r=-1$ does not lead to an independent solution, but points us into the right direction. Noting that the derivative of $\ln \frac{1-u}u$ is given by $-1/u(1-u)$, a term that cancels the polynomial occurring under the second derivative of equation (\ref{eq_flow_lin}), we insert the following ansatz
\begin{equation}
\label{eq_log_sol}
f_a(u) = \frac{q(u)}{u(1-u)} + p(u)\ln \frac{1-u}u
\end{equation}
with $p(u),q(u)$ polynomials in $u$. Equating terms in $\ln\frac{1-u}u$ as well as terms in $1/u(1-u)$ leads to the following two equations:
\begin{subequations}
\begin{eqnarray}
  0 &=& 4(1-3a/\epsilon) p(u) + 2(1-2u)p'(u) \\
 && {} + u(1-u)p''(u), \nonumber \\
  0 &=& 6(1-2a/\epsilon) q(u) + u(1-u) q''(u) \\
 && {} - (1-2u)p(u) - 2u(1-u)p'(u). \nonumber
\end{eqnarray}
\end{subequations}
$p(u)$ obeys the same equation as $f_a(u)$, hence the second type of logarithmic solutions has the same set of eigenvalues as the first type (eq.~(\ref{eq_evalues})), and these eigenvalues (including $a=\epsilon/2$) are the only ones possible.

$p(u)$ gives us $q(u)$, again by comparing coefficients, and thus we have the complete set of finite-series solutions. The first few are listed in table \ref{tab_eigenfunctions}.
\begin{table}[b]
\begin{center}
\hspace{1cm}
\begin{tabular}{c|c|c c}
$m$ & $\tilde a_m / \epsilon$ & $f_m^{even}(u)$ & $f_m^{odd}(u)$ \\
\hline
$-1$ & $\frac 1 2$  &  $\frac{c_e}{u(1-u)}$  &  $\frac{c_o u}{u(1-u)}$ \\
$0$ & $\frac 1 3$  &  $c_e$  &  $c_o \left( \frac{1-2u}{u(1-u)} +2 \ln{\frac{1-u}u}\right)$ \\
$1$ & $0$  &  $c_e \left( \frac {1-12u(1-u)}{u(1-u)} + 6(1-2u)\ln{\frac{1-u}u}\right)$  &  $c_o(1-2u)$ \\
$2$ & $-\frac 1 2$  &  $c_e(1-5u(1-u))$  &  $c_o\left( \frac{(1-2u)(1-30u(1-u))}{u(1-u)} - 12(1-5u(1-u))\ln{\frac{1-u}u}\right)$
\end{tabular}
\caption{First few eigenfunctions of the linearized flow equation (\ref{eq_flow_lin}) at zero temperature.}
\label{tab_eigenfunctions}
\end{center}
\end{table}
None of the finite-series eigenfunctions, however, fulfill the zero-mean condition $\int_0^1 \dif u f_a(u) = 0$. From equation (\ref{eq_flow_lin}) one has
\begin{equation}
\int_0^1 \dif u f_a(u) = - \frac 1 {6(1-2a/\epsilon)} \int_0^1 \dif u [ u(1-u) f_a(u)]''.
\end{equation}
In the case of the first type of solutions
\begin{equation}
\int_0^1 \dif u f_a(u) = \frac 1 {6(1-2a/\epsilon)} [ f_a(0) + f_a(1) ],
\end{equation}
which is non-zero for the even solutions we are interested in.

In the case of the logarithmic solutions (eq.~(\ref{eq_log_sol})):
\begin{eqnarray}
&\int_0^1 \dif u f_a(u) =& - \frac 1 {6(1-2a/\epsilon)} \int_0^1 \dif u \bigg[ q(u) \\
 && {} + u(1-u) \log\frac{1-u}u p(u)\bigg]'' \nonumber \\
&= \frac 1 {6(1-2a/\epsilon)} \big[ &q'(0) - q'(1) + p(1) - p(0)\big],
\end{eqnarray}
which is not zero either, for the even solutions.

\pagebreak
\section{Thermal boundary layer}
\label{app:tbl}

We now insert the ansatz of eq.~(\ref{ansatztbl}) into the FRG equations,
for $\partial_l (\Delta(u) - \Delta(0))$ and $\partial_l \Delta(0)$
respectively, collecting orders in $O(T_l)$, and using $\partial_l
\equiv - \theta T_l
\partial_{T_l}$ one finds to first order:
\begin{eqnarray}
&& 0 =  \phi_1''(0) - \phi_1''(x) - \frac{1}{2} (\phi_1(x)^2)'' \\
&& 0 = (\epsilon - 2 \zeta) \Delta^*(0) - \chi^2 \phi_1''(0) = 0
\end{eqnarray}
The solution is given in the text. The second order in $\tilde T_l$ yields:
\begin{eqnarray}
 && (\theta - \zeta) x \phi'_1(x) - (\theta + \epsilon - 2 \zeta)
\phi_1(x) = \\ 
&& \hspace{2cm} \chi^2 ( (\phi_1(x) \phi_2(x))'' + \phi_2''(x) -
\phi_2''(0)) \quad\quad \nonumber \\
 && - (\theta + \epsilon - 2 \zeta) \gamma_1 = \chi^2 \phi_2''(0)
\label{phi2}
\end{eqnarray}
whose solution is:
\begin{eqnarray}
 \phi_2(x) &=& \frac 1 {\phi(x)} \bigg[  \frac{1}{2} \phi_2''(0) x^2  \\
  && {} + \frac{1}{\chi^2} \bigg( (12 \epsilon - 48 \zeta + 36 \theta) (\phi(x)-1) \nonumber\\
  && {} + \frac{1}{2} x^2 (\theta + \epsilon - 2 \zeta) + \frac{1}{6} x^2 \phi(x) (\zeta - \epsilon) \nonumber \\
  && {} - 3(\epsilon-3 \zeta +2\theta)\,x\,\text{ArcSinh}(\frac{x}{6}) \bigg) \bigg] \nonumber
\end{eqnarray}
Fixing the free parameter $\phi_2''(0)$ by means of eq.~(\ref{phi2}), the
above simplifies into:
\begin{eqnarray}
 \phi_2(x) &=& \frac 1 {\chi^2 \phi(x)} \bigg[ 12 (\epsilon - 4 \zeta + 3
\theta) (\phi(x)-1) \\
 && {} + \frac{1}{2} x^2 (\theta + \epsilon - 2 \zeta) (1-\gamma_1)
       + \frac{1}{6} x^2 \phi(x) (\zeta - \epsilon) \nonumber \\
 && {} - 3(\epsilon-3 \zeta +2\theta)\,x\,\text{ArcSinh}(\frac{x}{6}) \bigg] \nonumber
\end{eqnarray}
This can be further simplified using $\theta=2-\epsilon+2\zeta$
to yield the formula given in the text. One can check that this agrees with the result
of \cite{ChauveBL} once corrected for a misprint ($f_3(x) \to
f_3(x)+x^2 (\epsilon-\zeta)/3 $ in the formula given in Appendix D3
of \cite{ChauveBL}). There the TBL expansion of the quantity
$y(u)=(\Delta(u)-\Delta(0)-T)^2=T^2 f_2(\chi u/6) + \chi^{-2} T^3
f_3(\chi u/6) +..$ was computed.


\begin{thebibliography}{10}
\bibitem{subdominant} This is the usual situation but there can be exceptions such as the case of subdominant chaos, where the perturbation is of shorter range than the original
disorder, e.g. $V_i=V + \delta W$ where $V$ is RF and $W$ is RB.
\end{thebibliography}
\end{document}